\documentstyle[12pt]{article}
\def\be{\begin{equation}}
\def\ee{\end{equation}}
\def\bea{\begin{eqnarray}}
\def\eea{\end{eqnarray}}

\begin{document}

\begin{flushright}
IPM/P-2001/001\\
hep-th/0104089
\end{flushright}

\begin{center}
{\Large{\bf General Target Space Duality and Its Effects on D-branes }}                  
										 
\vskip .5cm   
{\large Davoud  Kamani}
\vskip .1cm
 {\it Institute for Studies in Theoretical Physics and
Mathematics
\\  P.O.Box: 19395-5531, Tehran, Iran}\\
{\sl E-mail: kamani@theory.ipm.ac.ir}
\\
\end{center}

\begin{abstract} 

We study a special combination of the compact directions
of the spacetime (compactified on tori)
and the worldsheet parity transformed of these directions.
The transformations that change the compact part of the spacetime to this 
combination, are more general than the T-duality transformations. 
Many properties of this combination and also 
of the corresponding worldsheet fermions
are studied. By using the boundary state formalism, 
we study the effects of the above transformations to D-branes.
For the special cases the resulted branes 
reduce to the known mixed branes, or reduce to the modified mixed branes.

\end{abstract} 
\vskip .5cm


\newpage
\section{Introduction}
 
T-duality transformation is an exact symmetry of the closed string theory
\cite{1}. The problems that might at first sight seem different, can be
related by T-duality. In type II superstring theory, 
the action of the T-duality
on a compact direction of a D-brane, changes this direction to a transverse 
direction of a new brane. When T-duality acts 
on a transverse compact direction of 
a D-brane, this direction becomes along the new 
brane. Therefore the action of 
T-duality on the odd number of the compact directions changes the type
IIA theory to the type IIB theory and vice-versa \cite{2,3,4}. This means, 
the type IIA theory compactified 
on a circle of radius R is equivalent to the type
IIB theory compactified on a circle of radius $\alpha'/R$ \cite{1,2,5,6,7}.

We concentrate on the compact part of 
the spacetime (compactified on tori) and its 
worldsheet parity transformed. 
In fact, by special combination of these parts (Y-space),
the target-space duality is generalized. 
In this combined space the mass operator of the 
closed superstring is the same as that in the spacetime. 
Application of the transformations
that transform spacetime to the above combination to a D-brane, produces a 
brane that is more general than a mixed brane. 
For the special cases of these transformations the super 
virasoro operators and consequently the Hamiltonian of the
closed superstring are invariant.
Note that the boundary conditions of the closed superstring, emitted
by a mixed brane, are combinations of Dirichlet and Neumann boundary
conditions. This is due to the NS$\otimes$NS background B-field 
\cite{2,8,9,10,11}. 

This paper is organized as the follows. 
In section 2, we give a brief review of the
T-duality and its effects on D-branes. In section 3, some 
properties of the Y-space will be studied. 
This space appears similar to a compact space, therefore
under the duality transformations, 
some interesting effects appear. In section 4,
the corresponding fermions of the $Y$-space will be obtained. In section 5,
we apply the transformations that change spacetime to the Y-space, 
to the D-branes. We obtain the transformed  
boundary state of a closed superstring, emitted from a brane. 
In section 6, by neglecting the worldsheet parity 
part of the $Y$-space we obtain the usual mixed branes.
Putting away the spacetime part of it, leads to the modified mixed branes.
The field strengths on these branes, are some combinations of the
transformation matrices that change spacetime to the $Y$-space. 

A useful tool for describing branes (specially 
in non-zero background fields) is boundary state formalism
\cite{12,13,14,15}. In studying the branes, we will use this formalism.
\section{T-duality} 
 
Since we will use the T-duality in the next sections, 
we discuss the T-duality in type II superstring theory 
in this section. The mode expansion
of the closed string coordinate $X^\mu$, is
\bea
X^\mu (\sigma , \tau) = X^\mu _L (\tau+\sigma)   
+X^\mu _R (\tau-\sigma) \;\;,
\eea
\bea
X^\mu _L (\tau + \sigma)=x^\mu_L +2\alpha'p^\mu_L   
(\tau+\sigma) +\frac{i}{2} \sqrt{2\alpha'} \sum_{n \neq 0} \bigg{(}
\frac{1}{n}
{\tilde{\alpha}}^{\mu}_n e^{-2in(\tau+\sigma)} \bigg{)}\;\;,
\eea
\bea
X^\mu _R ( \tau-\sigma )= x^\mu_R +2\alpha'p^\mu_R   
(\tau-\sigma) +\frac{i}{2} \sqrt{2\alpha'} \sum_{n \neq 0} \bigg{(}
\frac{1}{n}
{\alpha}^{\mu}_n e^{-2in(\tau-\sigma)} \bigg{)}\;\;.
\eea
For non-compact direction $X^\mu$, since $X^\mu$ must be single valued at 
$\sigma$ and $\sigma+\pi$, there is $p^\mu_L = p^\mu_R =\frac{1}{2}p^\mu$.

If the direction $X^{\bar \mu}$ is compactified on a circle of radius 
$R_{\bar \mu}$ , we have the identification 
\bea
 x^{\bar \mu} \equiv x^{\bar \mu}+2\pi L^{\bar \mu}\;\;,
 \eea
 where
\bea
 x^{\bar \mu}= x^{\bar \mu}_L+ x^{\bar \mu}_R\;\;,
 \eea
 \bea
 L^{\bar \mu}= N^{\bar \mu}R_{\bar \mu}\;\;,\;\;\;
 {\rm (no\;sum\;on}\;{\bar \mu})\;.  
 \eea
 Therefore the compacted coordinates
 at $\sigma+\pi$ and $\sigma$ satisfy
 \bea
 X^{\bar \mu}(\sigma+\pi , \tau) -  X^{\bar \mu}(\sigma , \tau) =
 2\pi L^{\bar \mu} ,
 \eea
 which gives
 \bea
 L^{\bar \mu}=\alpha'(p^{\bar \mu}_L- p^{\bar \mu}_R)=  
 N^{\bar \mu}R_{\bar \mu}\;\;,  
 \eea
 the integer $N^{\bar \mu}$ is the winding number of closed string around
the direction $ X^{\bar \mu}$. According 
  to the identification (4) the momentum
  component $p^{\bar \mu}$ of the closed string is quantized,
  \bea
p^{\bar \mu}= p^{\bar \mu}_L+ p^{\bar \mu}_R=  
 \frac{M^{\bar \mu}}{R_{\bar \mu}}\;\;,  
 \eea
where $M^{\bar \mu}$, the closed string momentum number, is an integer.

Under the T-duality transformations,
\bea
T_{\bar \mu} :\;\; \frac{R_{\bar \mu}}{\sqrt{\alpha'}} \leftrightarrow  
\frac{\sqrt{\alpha'}}{R_{\bar \mu}} \;\;\;\;,\;\;\;\;
M^{\bar \mu} \leftrightarrow N^{\bar \mu} \;, 
\eea
the closed string mass operator 
and the virasoro operators are invariant. According to 
the eqs. (8)-(10) there are $p^{\bar \mu}_L \rightarrow p^{\bar \mu}_L$ and 
$p^{\bar \mu}_R \rightarrow -p^{\bar \mu}_R$. Generalizing this for all
oscillators i.e.
\bea
\left \{ \begin{array}{cl}
{\tilde \alpha}^{\bar \mu}_n \rightarrow {\tilde \alpha}^{\bar \mu}_n \;,\\
\alpha^{\bar \mu}_n \rightarrow -\alpha^{\bar \mu}_n \;, \\ 
\end{array} \right.
\eea
and also considering
\bea
\left \{ \begin{array}{cl}
x^{\bar \mu}_L \rightarrow x^{\bar \mu}_L \;\;\;, \\ 
x^{\bar \mu}_R \rightarrow -x^{\bar \mu}_R \;\;\;,
\end{array} \right.
\eea
we obtain
\bea
T_{\bar \mu}\;:\; X^{\bar \mu}(\sigma , \tau) \rightarrow
 X'^{\bar \mu}(\sigma , \tau) = X^{\bar \mu}_L - X^{\bar \mu}_R \;\;,
 \eea
 which is spacetime parity transformation acting only on the right moving
 degree of freedom.
 
 According to the eqs. (2) and (3), the dual coordinate 
 $X'^{\bar \mu}(\sigma , \tau)$ at $\sigma$ and $\sigma+\pi$ is not single
 valued, 
 \bea
 X'^{\bar \mu}(\sigma+\pi , \tau) -  X'^{\bar \mu}(\sigma , \tau) =
 2\pi \alpha' p^{\bar \mu} =2\pi M^{\bar \mu} \frac{\alpha'}
 {R_{\bar \mu}}\;\;,
 \eea
 which says the dual coordinate $X'^{\bar \mu}$ is compact  on a circle with
 radius $R'_{\bar \mu}=\alpha'/R_{\bar \mu}$ . 
 Therefore we have the following identification
\bea
 x'^{\bar \mu} \equiv x'^{\bar \mu}+2\pi \alpha'p^{\bar \mu}\;,
 \eea
 where
\bea
 x'^{\bar \mu}= x^{\bar \mu}_L- x^{\bar \mu}_R\;.
 \eea

Substitution of eqs. (2) and (3) in the transformation
(13) gives the exchange
\bea
\partial_{\tau}X^{\bar \mu} \leftrightarrow \partial_\sigma 
X^{\bar \mu}\;,
\eea
i.e. under the T-duality, the Dirichlet and Neumann boundary 
conditions of the closed string exchange.
Worldsheet supersymmetry requires the following T-duality transformations 
on the worldsheet fermions
\bea
T_{\bar \mu}\;:\;
\left \{ \begin{array}{cl}
\psi^{\bar \mu}_+ \rightarrow \psi^{\bar \mu}_+ \;\;\;,\\
\psi^{\bar \mu}_- \rightarrow -\psi^{\bar \mu}_-\;\;\;.
\end{array} \right.
\eea
Now we apply the T-duality transformations 
on a D$_p$-brane, and briefly study their 
effects on branes.

{\bf T-duality and D-branes}

A D$_p$-brane can be described by the boundary state $\mid B \rangle$,
\cite{12,13,14,15}.
This state satisfies the following boundary conditions of the closed string 
emitted by a D$_p$-brane 
\bea
(\partial_\tau X^\alpha)_{\tau=0} | B \rangle =0 \;,
\eea
\bea
(\partial_\sigma X^i)_{\tau=0} | B \rangle =0 \;,
\eea
where the set $\{ X^\alpha \}= 
\{ X^0 ,  X^{\alpha_1} ,...,X^{\alpha_p} \}$ shows the directions along the
world volume of the brane and the set $\{X^i\}$ shows
the directions perpendicular to the D$_p$-brane. The fermionic part of the
closed string boundary conditions are
\bea
(\psi^\alpha_- -i\eta \psi^\alpha_+)_{\tau=0} | B \rangle =0 \;,
\eea
\bea
(\psi^i_- +i\eta \psi^i_+)_{\tau=0} | B \rangle =0 \;,
\eea
where $\eta=\pm1$ is introduced to simplify the ``GSO'' projection.

Now assume that the $X^{\bar \alpha}$-direction of the above D$_p$-brane is
compact on a circle. The action of the T-duality (i.e. transformations
(17) and (18)) on this direction changes the boundary conditions (19) and
(21) for $\alpha ={\bar \alpha}$, i.e.
\bea
(\partial_\sigma X^{\bar \alpha})_{\tau=0} | B' \rangle =0 \;,
\eea
\bea
(\psi^{\bar \alpha}_- +i\eta 
\psi^{\bar \alpha}_+)_{\tau=0} | B' \rangle =0 \;.
\eea
This means that the direction $X^{\bar \alpha}$ is perpendicular to the new 
brane, which $\mid B' \rangle$ describes it. 
Hence one unit of the dimension of the brane is reduced, which
means for even (odd) values of $p$ type IIA 
(type IIB) theory changes to type IIB (type IIA) theory.

The action of the T-duality on the transverse compact 
direction $X^{\bar i}$,
changes the equations (20) and (22) for $i={\bar i}$, as
\bea
(\partial_\tau X^{\bar i})_{\tau=0} | B'' \rangle =0 \;,
\eea
\bea
(\psi^{\bar i}_- -i\eta \psi^{\bar i}_+)_{\tau=0} | B'' \rangle =0 \;\;,
\eea
therefore the direction $X^{\bar i}$ is along the brane. The corresponding
boundary state of this brane is $| B'' \rangle$.

Application of T-duality on arbitrary number of brane directions
and transverse directions to it, 
changes the brane to another brane, that can 
be obtained by the above method.
One can obtain a brane with background fields 
from a D$_p$-brane by appropriate actions of T-duality. For example the
action of the T-duality along the directions that make an 
angle with the brane, 
or along the boosted directions of the brane produces a brane 
with magnetic and electric fields \cite{10,11}.

\section{The Combined Space $\{ Y^{\bar \mu} \}$}

{\bf The Subspace $\{ {\tilde X}^{\bar \mu} \}$}

Now we define the tilde coordinate ${\tilde X}^{\bar \mu} 
={\tilde X}^{\bar \mu}_L+{\tilde X}^{\bar \mu}_R$ as follows
\bea
{\tilde X}^{\bar \mu}_L (\tau + \sigma)={\tilde x}^{\bar \mu}_L
+2\alpha'{\tilde p}^{\bar \mu}_L   
(\tau+\sigma) +\frac{i}{2} \sqrt{2\alpha'} \sum_{n \neq 0} 
\bigg{(}\frac{1}{n}
{\alpha}^{\bar \mu}_n e^{-2in(\tau+\sigma)} \bigg{)}\;,
\eea
\bea
{\tilde X}^{\bar \mu}_R (\tau - \sigma)={\tilde x}^{\bar \mu}_R 
+2\alpha'{\tilde p}^{\bar \mu}_R
(\tau-\sigma) +\frac{i}{2} \sqrt{2\alpha'} \sum_{n \neq 0} 
\bigg{(}\frac{1}{n}
{\tilde \alpha}^{\bar \mu}_n e^{-2in(\tau-\sigma)} \bigg{)}\;,
\eea
where
\bea
\left \{ \begin{array}{cl}
{\tilde p}^{\bar \mu}_L = \frac{1}{\sqrt{2\alpha'}}\alpha^{\bar \mu}_0 
=p^{\bar \mu}_R \;, \\ 
{\tilde p}^{\bar \mu}_R = \frac{1}{\sqrt{2\alpha'}}
{\tilde \alpha}^{\bar \mu}_0 =p^{\bar \mu}_L \;, \\ 
\end{array} \right.
\eea
which gives ${\tilde p}^{\bar \mu}= p^{\bar \mu}$.
For appropriate ${\tilde x}^{\bar \mu}_R$ and 
${\tilde x}^{\bar \mu}_L$, for example 
\bea
\left \{ \begin{array}{cl}
{\tilde x}^{\bar \mu}_L = x^{\bar \mu}_R\;\;\;, \\ 
{\tilde x}^{\bar \mu}_R = x^{\bar \mu}_L\;\;\;, 
\end{array} \right.
\eea
the coordinate ${\tilde X}^{\bar \mu}(\sigma , \tau)$ is the worldsheet 
parity transformed $(\sigma \rightarrow -\sigma )$ 
of $X^{\bar \mu}(\sigma , \tau)$, that is
\bea
{\tilde X}^{\bar \mu}(\sigma , \tau)=X^{\bar \mu}(-\sigma , \tau)\;\;.
\eea

If the subspace $\{ X^{\bar \mu} \}$ is compact, 
according to the equation (29), the tilde subspace will be also compact,
\bea
{\tilde L}^{\bar \mu} = \alpha'({\tilde p}^{\bar \mu}_L - 
{\tilde p}^{\bar \mu}_R) = - L^{\bar \mu}\;,
\eea
which means, worldsheet parity changes the windings of the closed string. 
There is therefore an identification,
\bea
{\tilde x}^{\bar \mu} \equiv {\tilde x}^{\bar \mu} +2\pi {\tilde L} 
^{\bar \mu}= {\tilde x}^{\bar \mu} -2\pi L^{\bar \mu}\;\;. 
\eea
The dual coordinate of the ${\tilde X}^{\bar \mu}$ is
\bea
{\tilde X}'^{\bar \mu} (\sigma , \tau) =
{\tilde X}^{\bar \mu}_L (\tau+\sigma)  
-{\tilde X}^{\bar \mu}_R (\tau-\sigma) \;.
\eea
Since ${\tilde X}'^{\bar \mu}$ is a compact coordinate, 
${\tilde {x'}}^{\bar \mu}=
{\tilde x}^{\bar \mu}_L-{\tilde x}^{\bar \mu}_R$
is identified with ${\tilde {x'}}^{\bar \mu}+2\pi \alpha' 
{\tilde p}^{\bar \mu}$,
\bea            
{\tilde {x'}}^{\bar \mu} \equiv  {\tilde {x'}}^{\bar \mu}+2\pi \alpha' 
{\tilde p}^{\bar \mu} =  {\tilde {x'}}^{\bar \mu}
+2\pi \alpha' p^{\bar \mu}\;.
\eea


{\bf The Y-space}

This space is the following 
combination of the subspaces $\{X^{\bar \mu}\}$
and $\{{\tilde X}^{\bar \mu} \}$. The left and the right moving parts of
the coordinates of this combined space are defined by
\bea
\left \{ \begin{array}{cl}
Y^{\bar \mu}_R  = {\cal{A}}^{\bar \mu} _{\;\;\; {\bar \nu}} 
X^{\bar \nu}_R (\tau -\sigma) + 
{\tilde {\cal{A}}}^{\bar \mu}_{\;\;\;{\bar \nu}}
{\tilde X}^{\bar \nu}_R (\tau-\sigma)\;\;\;,\\
Y^{\bar \mu}_L  = {\cal{B}}^{\bar \mu}_{\;\;\;{\bar \nu}}
{\tilde X}^{\bar \nu}_L (\tau +\sigma) + 
{\tilde {\cal{B}}}^{\bar \mu}_{\;\;\;{\bar \nu}}
 X^{\bar \nu}_L (\tau+\sigma)\;\;\;,\\
\end{array} \right.
\eea
therefore
\bea
Y^{\bar \mu}(\sigma , \tau) = Y^{\bar \mu}_L (\tau+ \sigma)
+Y^{\bar \mu}_R (\tau- \sigma)\;.
\eea
The matrices ${\cal{A}}$, ${\tilde{\cal{A}}}$, ${\cal{B}}$ and
${\tilde {\cal{B}}}$ will be limited by some conditions such as  the 
invariance of the mass operator of closed string, 
under changing $\{X^{\bar \mu}\}$-subspace to the $Y$-space.
As next we will discuss about their properties.
Introducing mode expansions (2), (3), (27) and (28) in the definition (36), 
shows the oscillating modes of the $Y^{\bar \mu}$
\bea
Y^{\bar \mu}_L =y^{\bar \mu}_L 
+2\alpha' {\Pi}^{\bar \mu}_L   
(\tau+\sigma) +\frac{i}{2} 
\sqrt{2\alpha'} \sum_{n \neq 0} \bigg{(} \frac{1}{n}
{\tilde a}^{\bar \mu}_n e^{-2in(\tau+\sigma)} \bigg{)}\;,
\eea
\bea
Y^{\bar \mu}_R=y^{\bar \mu}_R 
+2\alpha' {\Pi}^{\bar \mu}_R   
(\tau-\sigma) +\frac{i}{2} 
\sqrt{2\alpha'} \sum_{n \neq 0} \bigg{(}\frac{1}{n}
a^{\bar \mu}_n e^{-2in(\tau-\sigma)} \bigg{)}\;,
\eea
where the parameters are
\bea
y^{\bar \mu}_R = {\cal{A}}^{\bar \mu} _{\;\;\; {\bar \nu}} 
x^{\bar \nu}_R + {\tilde {\cal{A}}}^{\bar \mu}_{\;\;\;{\bar \nu}}
{\tilde x}^{\bar \nu}_R \;,
\eea
\bea
y^{\bar \mu}_L = {\cal{B}}^{\bar \mu} _{\;\;\; {\bar \nu}} 
{\tilde x}^{\bar \nu}_L + {\tilde {\cal{B}}}^{\bar \mu}_{\;\;\;{\bar \nu}}
x^{\bar \nu}_L \;,
\eea
\bea
\Pi^{\bar \mu}_R = {\cal{A}}^{\bar \mu} _{\;\;\; {\bar \nu}} 
p^{\bar \nu}_R + {\tilde {\cal{A}}}^{\bar \mu}_{\;\;\;{\bar \nu}}
p^{\bar \nu}_L \;,
\eea
\bea
\Pi^{\bar \mu}_L = {\cal{B}}^{\bar \mu} _{\;\;\; {\bar \nu}} 
p^{\bar \nu}_R + {\tilde {\cal{B}}}^{\bar \mu}_{\;\;\;{\bar \nu}}
p^{\bar \nu}_L \;,
\eea
\bea
a^{\bar \mu}_n = {\cal{A}}^{\bar \mu} _{\;\;\; {\bar \nu}} 
\alpha^{\bar \nu}_n + {\tilde {\cal{A}}}^{\bar \mu}_{\;\;\;{\bar \nu}}
{\tilde \alpha}^{\bar \nu}_n \;,
\eea
\bea
{\tilde a}^{\bar \mu}_n = {\cal{B}}^{\bar \mu} _{\;\;\; {\bar \nu}} 
\alpha^{\bar \nu}_n + {\tilde {\cal{B}}}^{\bar \mu}_{\;\;\;{\bar \nu}}
{\tilde \alpha}^{\bar \nu}_n \;.
\eea
To find eqs. (42) and (43) we have used the eq. (29).

Now look at the mass operator of the closed string. The bosonic part of it, 
is 
\bea
\alpha' M^2_b&=& \alpha^{\bar \mu}_0 \alpha_{0 {\bar \mu}}  
+{\tilde \alpha}^{\bar \mu}_0 {\tilde \alpha}_{0 {\bar \mu}}
+2\sum_{n=1}^{\infty}(\alpha^{\bar \mu}_{-n} \alpha_{n {\bar \mu}}  
+{\tilde \alpha}^{\bar \mu}_{-n} {\tilde \alpha}_{n {\bar \mu}}) \nonumber\\
&~&+2\sum_{n=1}^{\infty}(\alpha^{ \mu'}_{-n} \alpha_{n { \mu'}}  
+{\tilde \alpha}^{\mu'}_{-n} {\tilde \alpha}_{n {\mu'}})
+{\rm ghost\;part} \;,
\eea
where $\mu' \notin \{{\bar \mu}\}$. Consider the transformations
\bea
\left \{ \begin{array}{cl}
X^{\bar \mu}_L \rightarrow Y^{\bar \mu}_L \;, \\
X^{\bar \mu}_R \rightarrow Y^{\bar \mu}_R \;,  
\end{array} \right.
\eea
which give $\alpha^{\bar \mu}_n \rightarrow a^{\bar \mu}_n$ and
${\tilde \alpha}^{\bar \mu}_n \rightarrow {\tilde a}^{\bar \mu}_n$.
The invariance of the mass operator under the transformations (47)
gives the following conditions on the matrices 
 ${\cal{A}}$, ${\tilde{\cal{A}}}$, ${\cal{B}}$ and
${\tilde {\cal{B}}}$,  
\bea
{\cal{A}}^T{\cal{A}} + {\cal{B}}^T {\cal{B}}={\bf 1} \;,
\eea
\bea
{\tilde {\cal{A}}}^T{\tilde {\cal{A}}} 
+ {\tilde {\cal{B}}}^T {\tilde {\cal{B}}}={\bf 1} \;,
\eea
\bea
{\cal{A}}^T{\tilde {\cal{A}}} 
+ {\cal{B}}^T {\tilde {\cal{B}}}={\bf 0} \;.
\eea
One can show that these matrices satisfy 
the following relations 
\bea
{\cal{A}}{\cal{A}}^T +{\tilde {\cal{A}}} {\tilde {\cal{A}}}^T = {\bf 1}\;,
\eea
\bea
{\cal{B}}{\cal{B}}^T +{\tilde {\cal{B}}} {\tilde {\cal{B}}}^T = {\bf 1}\;,
\eea
\bea
{\cal{A}}{\cal{B}}^T +{\tilde {\cal{A}}} {\tilde {\cal{B}}}^T = {\bf 0}\;.
\eea
These equations are not independent of eqs.
(48)-(50). We will use them for later purposes.

For some special choices of these matrices, the $Y$-space reduces to the 
known cases. For example the $Y$-space is,

$\{X^{\bar \mu}\}$-subspace for,
${\tilde {\cal{A}}} ={\cal{B}}={\bf 0}$, 
${\cal{A}} ={\tilde {\cal{B}}}={\bf 1}$,

T-dual of $\{X^{\bar \mu}\}$-subspace for,
${\tilde {\cal{A}}} ={\cal{B}}={\bf 0}$, 
${\cal{A}} =-{\tilde {\cal{B}}}=-{\bf 1}$,

$\{{\tilde X}^{\bar \mu}\}$-subspace for,
${\cal{A}} ={\tilde {\cal{B}}}={\bf 0}$, 
${\tilde {\cal{A}}} = {\cal{B}}={\bf 1}$,

T-dual of $\{X^{\bar \mu}\}$-subspace for,
$ {\cal{A}} ={\tilde{\cal{B}}}={\bf 0}$, ${\tilde {\cal{A}}} 
=-{\cal{B}}=-{\bf 1}$.

These choices of the matrices satisfy all conditions (48)-(53).

Oscillators of the $Y$-space, $a^{\bar \mu}_n $ 
and ${\tilde a}^{\bar \mu}_n $
satisfy the same commutation relations as $\alpha ^{\bar \mu}_n$ and
${\tilde \alpha}^{\bar \mu}_n$,
\bea
\left \{ \begin{array}{cl}
[a^{\bar \mu}_m , a^{\bar \nu}_n] = [{\tilde a}^{\bar \mu}_m , 
{\tilde a}^{\bar \nu}_n] = m \delta_{m+n,0} 
\eta^{{\bar \mu} {\bar \nu}} \;,\\
\hspace{-3.5cm {[ a^{\bar \mu}_m , {\tilde a}^{\bar \nu}_n ]} =0 }\;.
\end{array} \right.
\eea
Also $y^{\bar \mu}$ and $\Pi^{\bar \mu}$ are
\bea
y^{\bar \mu} = y^{\bar \mu}_R+y^{\bar \mu}_L  
= {\cal{A}}^{\bar \mu} _{\;\;\; {\bar \nu}} 
x^{\bar \nu}_R + {\tilde {\cal{A}}}^{\bar \mu}_{\;\;\;{\bar \nu}} 
{\tilde x}^{\bar \nu}_R+ 
{\cal{B}}^{\bar \mu} _{\;\;\; {\bar \nu}} 
{\tilde x}^{\bar \nu}_L + {\tilde {\cal{B}}}^{\bar \mu}_{\;\;\;{\bar \nu}} 
x^{\bar \nu}_L \;,
\eea
\bea
\Pi^{\bar \mu}=\Pi^{\bar \mu}_L+\Pi^{\bar \mu}_R
= \frac{1}{2}( {\cal{A}}+{\cal{B}}+{\tilde {\cal{A}}}+{\tilde {\cal{B}}}) 
^{\bar \mu}_{\;\;\; {\bar \nu}}p^{\bar \nu} + \frac{1}{2\alpha'}
( -{\cal{A}}-{\cal{B}}+{\tilde {\cal{A}}}+{\tilde {\cal{B}}}) 
^{\bar \mu}_{\;\;\; {\bar \nu}}L^{\bar \nu} \;, 
\eea
They satisfy the following commutation relation,
\bea
[y^{\bar \mu} , \Pi^{\bar \nu}] = i \eta^{{\bar \mu}{\bar \nu}} \;\;\;,
\eea
where we have used the relation (30).

Define $\frac{1}{\alpha'} \Lambda^{\bar \mu}$ 
as the difference of $\Pi^{\bar \mu}_L$ and $\Pi^{\bar \mu}_R$
\bea
\frac{1}{\alpha'} \Lambda^{\bar \mu} =\Pi^{\bar \mu}_L-\Pi^{\bar \mu}_R
= \frac{1}{2}( -{\cal{A}}+{\cal{B}}-{\tilde {\cal{A}}}+{\tilde {\cal{B}}}) 
^{\bar \mu}_{\;\;\; {\bar \nu}}p^{\bar \nu} + \frac{1}{2\alpha'}
({\cal{A}}-{\cal{B}}-{\tilde {\cal{A}}}+{\tilde 
{\cal{B}}})^{\bar \mu}_{\;\;\;{\bar \nu}} L^{\bar \nu}\;\;. 
\eea
We see that
\bea
Y^{\bar \mu}(\sigma+\pi , \tau)-Y^{\bar \mu}(\sigma , \tau)=2\pi
\Lambda^{\bar \mu}\;\;,
\eea
which means $Y^{\bar \mu}(\sigma , \tau)$ is not single valued at $\sigma$
and $\sigma+\pi$, therefore it appears like a compact coordinate.

{\bf The dual space $\{Y'^{\bar \mu}\}$}

Now we define the dual of the $Y$-space, i.e. $Y'$-space 
as the difference between
the left and the right moving parts of $Y^{\bar \mu}$
\bea
Y'^{\bar \mu}(\sigma , \tau) = Y^{\bar \mu}_L-Y^{\bar \mu}_R=
{\cal{B}}^{\bar \mu} _{\;\;\; {\bar \nu}} 
{\tilde X}^{\bar \nu}_L + {\tilde {\cal{B}}}^{\bar \mu}_
{\;\;\;{\bar \nu}} 
X^{\bar \nu}_L 
-{\cal{A}}^{\bar \mu}_{\;\;\;{\bar \nu}}X^{\bar \nu}_R
-{\tilde {\cal{A}}}^{\bar \mu}_{\;\;\;{\bar \nu}} {\tilde X}^{\bar \nu}_R\;. 
\eea
Similar to the $Y$-space, the $Y'$-space for some
special choices of the matrices reduces to the $\{X^{\bar \mu}\}$-subspace,
$\{{\tilde X}^{\bar \mu}\}$-subspace or 
to the duals of them. According to the
eq. (60) for dualizing, it is sufficient to change ${\cal{A}}$ and
${\tilde {\cal{A}}}$ as
\bea
\left \{ \begin{array}{cl}
{\cal{A}} \rightarrow -{\cal{A}} \;,\\
{\tilde {\cal{A}}} \rightarrow -{\tilde {\cal{A}}} \;.
\end{array} \right.
\eea

According to the eq. (30), $y'^{\bar \mu}$ of the dual space is
\bea
y'^{\bar \mu} = y^{\bar \mu}_L - y^{\bar \mu}_R
= -{\cal{A}}^{\bar \mu} _{\;\;\; {\bar \nu}} 
x^{\bar \nu}_R - {\tilde {\cal{A}}}^{\bar \mu}_{\;\;\;{\bar \nu}} 
{\tilde x}^{\bar \nu}_R 
+{\cal{B}}^{\bar \mu} _{\;\;\; {\bar \nu}} 
{\tilde x}^{\bar \nu}_L + {\tilde {\cal{B}}}^{\bar \mu}_{\;\;\;{\bar \nu}} 
x^{\bar \nu}_L\;. 
\eea
This with $\frac{1}{\alpha'} \Lambda^{\bar \mu}$ satisfy the relation
\bea
[ y'^{\bar \mu} , \frac{1}{\alpha'} \Lambda^{\bar \nu} ]=
i \eta^{{\bar \mu}{\bar \nu}} \;,
\eea
as expected. Under the dualizing of the $Y$-space, i.e. transformations (61),
there is the following exchange,
\bea
\Pi^{\bar \mu} \leftrightarrow \frac{1}{\alpha'} \Lambda^{\bar \mu}\;.
\eea
In studying the branes, we will see other effects of this dualizing.

Now return to the equations (4), (5), (15) and (16). We find the
following identifications for $x^{\bar \mu}_L$ and $x^{\bar \mu}_R$,
\bea
\left \{ \begin{array}{cl}
x^{\bar \mu}_L \equiv x^{\bar \mu}_L + \pi (L^{\bar \mu}+
\alpha' p^{\bar \mu}) \;,\\
x^{\bar \mu}_R \equiv x^{\bar \mu}_R + \pi (L^{\bar \mu}-
\alpha' p^{\bar \mu}) \;.
\end{array} \right.
\eea
Similarly the eqs. (33) and (35) give the identifications
\bea
\left \{ \begin{array}{cl}
{\tilde x}^{\bar \mu}_L \equiv {\tilde x}^{\bar \mu}_L + 
\pi (-L^{\bar \mu}+ \alpha' p^{\bar \mu}) \;,\\
{\tilde x}^{\bar \mu}_R \equiv 
{\tilde x}^{\bar \mu}_R - \pi (L^{\bar \mu}
+\alpha' p^{\bar \mu})\;.
\end{array} \right.
\eea
Therefore, there are following identifications for $y^{\bar \mu}$ 
and $y'^{\bar \mu}$,
\bea
y^{\bar \mu} \equiv y^{\bar \mu} +2\pi \Lambda^{\bar \mu} \;,
\eea
\bea
y'^{\bar \mu} \equiv y'^{\bar \mu} +2\pi \alpha' \Pi^{\bar \mu} \;.
\eea
Note that under dualizing, $y^{\bar \mu}$ and $y'^{\bar \mu}$  
exchange, therefore $\Pi^{\bar \mu}$ and 
$\frac{1}{\alpha'} \Lambda^{\bar \mu}$ must also be exchanged. 
This takes place as we saw in (64). Also the dual coordinate $Y'^{\bar
\mu}(\sigma , \tau)$ satisfies the equation,
\bea
Y'^{\bar \mu}(\sigma+\pi , \tau)-Y'^{\bar \mu}(\sigma , \tau)=2\pi
\alpha' \Pi^{\bar \mu}\;,
\eea
which is expected and is consistent with the identification (68).

Consider the transformations (47) from 
$\{X^{\bar \mu}\}$-subspace to $\{Y^{\bar \mu}\}$-space.
These transformations form a group 
that is isomorphic to the group $O(2k)$, where $k$ is the dimension of the
compact part of the spacetime $\{X^{\bar \mu}\}$. This
orthogonality implies that the squared mass of the  
closed string to be invariant.
\section{The fermions of Y-space}

Consider the R$\otimes$R and the NS$\otimes$NS sectors
of type II superstrings. 
In terms of the oscillating modes the worldsheet fermions of 
these sectors are
\bea
\left \{ \begin{array}{cl}
\psi^\mu_- = \sum\limits_{n \in Z}( d^\mu_n e^{-2in(\tau-\sigma)})\;, \\
\psi^\mu_+ = \sum\limits_{n \in Z}( {\tilde d}^\mu_n 
e^{-2in(\tau+\sigma)})\;,
\end{array} \right.
\eea
for the R$\otimes$R sector, and
\bea
\left \{ \begin{array}{cl}
\psi^\mu_- = \sum\limits_{r \in Z+\frac{1}{2}}( b^\mu_r 
e^{-2ir(\tau-\sigma)})\;, \\
\psi^\mu_+ = \sum\limits_{r \in Z+\frac{1}{2}}( {\tilde b}^\mu_r 
e^{-2ir(\tau+\sigma)} )\;,
\end{array} \right.
\eea
for the NS$\otimes$NS sector.             
According to the worldsheet supersymmetry, 
the tilde fermions ${\tilde \psi}
^{\bar \mu}_\pm $ corresponding to the 
$\{{\tilde X}^{\bar \mu}\}$-subspace, are 
\bea
\left \{ \begin{array}{cl}
{\tilde \psi}^{\bar \mu}_- = \sum\limits_{n \in Z}( {\tilde d}^{\bar \mu}_n 
e^{-2in(\tau-\sigma)})\;, \\
{\tilde \psi}^{\bar \mu}_+ = \sum\limits_{n \in Z}(  
d^{\bar \mu}_n e^{-2in(\tau+\sigma)})\;,
\end{array} \right.
\eea
for the R$\otimes$R sector, and
\bea
\left \{ \begin{array}{cl}
{\tilde \psi}^{\bar \mu}_- = \sum\limits_{r \in Z+\frac{1}{2}}
( {\tilde b}^{\bar \mu}_r 
e^{-2ir(\tau-\sigma)})\;, \\
{\tilde \psi}^{\bar \mu}_+ = \sum\limits_{r \in Z+\frac{1}{2}}( 
b^{\bar \mu}_r e^{-2ir(\tau+\sigma)} )\;,
\end{array} \right.
\eea
for the NS$\otimes$NS sector. 
The fermions (72) and (73), 
are worldsheet parity transformed ($\sigma \rightarrow -\sigma$) of the 
worldsheet fermions (70) and (71).

Let $\chi_- (\sigma , \tau)$ and $\chi_+ (\sigma , \tau)$ be the right 
and the left moving components of the 
fermions of the $Y$-space, therefore
\bea
\chi^{\bar \mu}_-(\tau-\sigma) = {\cal{A}}^{\bar \mu}_{\;\;\;{\bar \nu}}
\psi^{\bar \nu}_- + {\tilde {\cal{A}}}
^{\bar \mu}_{\;\;\;{\bar \nu}}{\tilde \psi}
^{\bar \nu}_- \;,
\eea
\bea
\chi^{\bar \mu}_+(\tau+\sigma) = {\cal{B}}^{\bar \mu}_{\;\;\;{\bar \nu}}
{\tilde \psi}^{\bar \nu}_+ + {\tilde {\cal{B}}}^{\bar \mu}_
{\;\;\;{\bar \nu}} \psi^{\bar \nu}_+ \;.
\eea
For the special choices of the matrices that  
mentioned for the $Y$-space, these
fermions reduce to the fermions of $\{X^{\bar \mu}\}$-subspace, 
$\{{\tilde X}^{\bar \mu}\}$-subspace, or to the fermions of their dual
spaces. Oscillators of the $Y$-space fermions are,
\bea
\left \{ \begin{array}{cl}
D^{\bar \mu}_n = {\cal{A}}^{\bar \mu}_{\;\;\;{\bar \nu}}   
d^{\bar \nu}_n+{\tilde {\cal{A}}}^{\bar \mu}_{\;\;\;{\bar \nu}}   
 {\tilde d}^{\bar \nu}_n\;, \\ 
{\tilde D}^{\bar \mu}_n = {\cal{B}}^{\bar \mu}_{\;\;\;{\bar \nu}}   
d^{\bar \nu}_n+{\tilde 
{\cal{B}}}^{\bar \mu}_{\;\;\;{\bar \nu}}   
{\tilde d}^{\bar \nu}_n\;,  
\end{array} \right.
\eea
for the R$\otimes$R sector. For the NS$\otimes$NS sector,  
$d^{\bar \mu}_n$ and ${\tilde d}^{\bar \mu}_n$ should be replaced with 
$b^{\bar \mu}_r$ and ${\tilde b}^{\bar \mu}_r$, respectively.
Oscillators $D^{\bar \mu}_n$ and ${\tilde D}^{\bar \mu}_n$ satisfy the same
anti-commutation relations as $d^{\bar \mu}_n$ and ${\tilde d}^{\bar \mu}_n$,
\bea
\left \{ \begin{array}{cl}
\{D^{\bar \mu}_m , D^{\bar \nu}_n\} = \{{\tilde D}^{\bar \mu}_m , 
{\tilde D}^{\bar \nu}_n\} =  \delta_{m+n,0} \eta^{{\bar \mu} 
{\bar \nu}} \;,\\
\hspace{-4cm {\{ D^{\bar \mu}_m , {\tilde D}^{\bar \nu}_n \}} = 0} \;.
\end{array} \right.
\eea
Similar relations hold for the NS$\otimes$NS sector.

For the dual space ${Y'}$, also there are corresponding fermions.
The right and the left moving components of these fermions are
\bea
\left \{ \begin{array}{cl}
\chi'^{\bar \mu}_- = -\chi^{\bar \mu}_- \;,\\
\chi'^{\bar \mu}_+ = \chi^{\bar \mu}_+ \;,
\end{array} \right.
\eea
where, we have used the transformations (61).

Again return to the mass operator of the closed string. In the 
R$\otimes$R sector, for instance its fermionic part has the form,
\bea
\alpha' M^2_f &=&
2\sum_{n=1}^{\infty}n(d^{\bar \mu}_{-n} d_{n {\bar \mu}}  
+{\tilde d}^{\bar \mu}_{-n} {\tilde d}_{n {\bar \mu}} )
+2\sum_{n=1}^{\infty}n(d^{ \mu'}_{-n} d_{n { \mu'}}  
+{\tilde d}^{\mu'}_{-n} {\tilde d}_{n {\mu'}}) \nonumber\\
&~&+{\rm superghost\;part} \;,
\eea
where $\mu' \notin \{{\bar \mu}\}$. Under the transformations
\bea
\left \{ \begin{array}{cl}
\psi^{\bar \mu}_-  \rightarrow \chi^{\bar \mu}_- \;,\\
\psi^{\bar \mu}_+ \rightarrow \chi^{\bar \mu}_+ \;,
\end{array} \right.
\eea
or equivalently
\bea
\left \{ \begin{array}{cl}
d^{\bar \mu}_n  \rightarrow D^{\bar \mu}_n \;,\\
{\tilde d}^{\bar \mu}_n \rightarrow {\tilde D}^{\bar \mu}_n \;,
\end{array} \right.
\eea
the fermionic part of the mass operator is also 
invariant. This invariance does not impose 
new conditions on the matrices ${\cal{A}}$,  ${\tilde {\cal{A}}}$,  
${\cal{B}}$ and ${\tilde {\cal{B}}}$. 
This also holds for the NS$\otimes$NS sector. 
\section{Transformed branes}

To study the effects of the transformations (47) and (80) 
on a D$_p$-brane, we discuss the transformations for the brane
directions and the transformations 
for the transverse directions of the brane.
For simplicity we neglect the transformations that mix coordinates
$\{ X^{\bar \alpha}_R \}$ with $\{ X^{\bar i}_R \}
\bigcup \{ {\tilde X}^{\bar i}_R \}$ and also
$\{ X^{\bar i}_R \}$ with  
$\{ X^{\bar \alpha}_R \} \bigcup \{ {\tilde X}^{\bar \alpha}_R \}$. 
Similarly for the left moving coordinates.
Consider the following transformations for the compact 
directions along the brane i.e. $\{X^{\bar \alpha}\}$,
\bea
\left \{ \begin{array}{cl}
X^{\bar \alpha }_R \rightarrow Y^{\bar \alpha}_R = A^{\bar \alpha} 
_{\;\;\;{\bar \beta}}X^{\bar \beta}_R + 
{\tilde A}^{\bar \alpha}_{\;\;\;{\bar \beta}} 
{\tilde X}^{\bar \beta}_R \;,\\
X^{\bar \alpha }_L \rightarrow Y^{\bar \alpha}_L = B^{\bar \alpha} 
_{\;\;\;{\bar \beta}}{\tilde X}^{\bar \beta}_L 
+ {\tilde B}^{\bar \alpha}_{\;\;\;{\bar \beta}} 
 X^{\bar \beta}_L \;,
\end{array} \right.
\eea
for the bosonic part, and
\bea
\left \{ \begin{array}{cl}
\psi^{\bar \alpha }_- \rightarrow \chi^{\bar \alpha}_- = A^{\bar \alpha} 
_{\;\;\;{\bar \beta}} \psi^{\bar \beta}_- + 
{\tilde A}^{\bar \alpha}_{\;\;\;{\bar \beta}} 
{\tilde \psi}^{\bar \beta}_- \;,\\
\psi^{\bar \alpha }_+ \rightarrow \chi^{\bar \alpha}_+ = B^{\bar \alpha} 
_{\;\;\;{\bar \beta}}{\tilde \psi}^{\bar \beta}_+ 
+ {\tilde B}^{\bar \alpha}_{\;\;\;{\bar \beta}} 
 \psi^{\bar \beta}_+ \;,
\end{array} \right.
\eea
for the fermionic part. The  four matrices $A$, ${\tilde A}$, $B$ and
${\tilde B}$ satisfy the conditions (48)-(53). For some of the compact 
transverse directions of the brane i.e. $\{ X^{\bar i}\}$, we introduce the 
transformations
\bea
\left \{ \begin{array}{cl}
X^{\bar i }_R \rightarrow Y^{\bar i}_R = A'^{\bar i} 
_{\;\;\;{\bar j}}X^{\bar j}_R + 
{\tilde A}'^{\bar i}_{\;\;\;{\bar j}} 
{\tilde X}^{\bar j}_R \;, \\
X^{\bar i}_L \rightarrow Y^{\bar i}_L = B'^{\bar i} 
_{\;\;\;{\bar j}}{\tilde X}^{\bar j}_L 
+ {\tilde B}'^{\bar i}_{\;\;\;{\bar j}} 
 X^{\bar j}_L \;,
\end{array} \right.
\eea
for the bosonic part, and
\bea
\left \{ \begin{array}{cl}
\psi^{\bar i}_- \rightarrow \chi^{\bar i}_- = A'^{\bar i} 
_{\;\;\;{\bar j}} \psi^{\bar j}_- + 
{\tilde A}'^{\bar i}_{\;\;\;{\bar j}} 
{\tilde \psi}^{\bar j}_- \;,\\
\psi^{\bar i}_+ \rightarrow \chi^{\bar i}_+ = B'^{\bar i} 
_{\;\;\;{\bar j}}{\tilde \psi}^{\bar j}_+ 
+ {\tilde B}'^{\bar i}_{\;\;\;{\bar j}} 
 \psi^{\bar j}_+ \;,
\end{array} \right.
\eea
for the fermionic part. Again four matrices $A'$, ${\tilde A}'$, $B'$ and
${\tilde B}'$ satisfy the conditions (48)-(53).

Now we apply the transformations (82)-(85) to the boundary state equations
corresponding to a D$_p$-brane, 
i.e. equations (19)-(22). For the bosonic part 
there are,
\bea                             
\left \{ \begin{array}{cl}
\hspace{-3cm\bigg{(} (A+{\tilde B})^{\bar \alpha}_{\;\;\;{\bar 
\beta}}\partial_\tau 
X^{\bar \beta} 
- (A-{\tilde B})^{\bar \alpha}_{\;\;\;{\bar \beta}}
\partial_\sigma X^{\bar \beta}}\\
+( {\tilde A}+B)^{\bar \alpha}_{\;\;\;{\bar \beta}}\partial_\tau 
{\tilde X}^{\bar \beta} - ({\tilde A}-B)^{\bar \alpha}_{\;\;\;{\bar \beta}}
\partial_\sigma {\tilde X}^{\bar \beta}\bigg{)}_{\tau=0} 
| B_t \rangle =0 \;,\\
\hspace{-5cm(\partial_\tau X^{\alpha'})_{\tau=0} | B_t \rangle =0 }\;,
\end{array} \right.
\eea
where the set $\{ X^{\alpha'} \}$ shows the 
non-compact directions along the brane. The state $\mid B_t \rangle $
is transformed boundary state that describes the new brane. Also the 
transformed form of the equation (20) is given by
\bea
\left \{ \begin{array}{cl}
\hspace{-3cm\bigg{(} (A'-{\tilde B'})^{\bar i}_{\;\;\;{\bar j}}\partial_\tau 
X^{\bar j} -  (A'+{\tilde B'})^{\bar i}_{\;\;\;{\bar j}}
\partial_\sigma X^{\bar j}}\\
+({\tilde A'}-B')^{\bar i}_{\;\;\;{\bar j}}\partial_\tau 
{\tilde X}^{\bar j} -  ({\tilde A'}+B')^{\bar i}_{\;\;\;{\bar j}}
\partial_\sigma {\tilde X}^{\bar j}\bigg{)}_{\tau=0} 
| B_t \rangle =0 \;,\\
\hspace{-6cm(\partial_\sigma X^{i'})_{\tau=0} | B_t \rangle =0 }\;,  
\end{array} \right.
\eea
where $X^{i'}$ is a non-compact direction transverse to the brane. 

Transformations on the fermionic equations
(21) and (22), give the following boundary state equations,
\bea
\left \{ \begin{array}{cl}
\bigg{(}A^{\bar \alpha}_{\;\;\;{\bar \beta}} \psi^{\bar \beta}_-  
-i\eta{\tilde B}^{\bar \alpha}_{\;\;\;{\bar \beta}}\psi^{\bar \beta}_+ 
+{\tilde A}^{\bar \alpha}_{\;\;\;{\bar \beta}} 
{\tilde \psi}^{\bar \beta}_- -i\eta B^{\bar \alpha }_{\;\;\;{\bar \beta}}
{\tilde \psi}^{\bar \beta}_+ \bigg{)}_{\tau=0} | B_t \rangle =0 \;,\\  
\hspace{-5cm(\psi^{\alpha'}_- -i\eta \psi^{\alpha'}_+)_{\tau=0} 
| B_t \rangle =0 }\;,
\end{array} \right.
\eea
\bea
\left \{ \begin{array}{cl}
\bigg{(} A'^{\bar i}_{\;\;\;{\bar j}} \psi^{\bar j}_- + 
i\eta{\tilde B}'^{\bar i}_{\;\;\;{\bar j}}\psi^{\bar j}_+ 
+{\tilde A}'^{\bar i}_{\;\;\;{\bar j}} 
{\tilde \psi}^{\bar j}_- +i\eta B'^{\bar i}_{\;\;\;{\bar j}}
{\tilde \psi}^{\bar j}_+ \bigg{)}_{\tau=0} | B_t \rangle =0 \;,\\  
\hspace{-5cm(\psi^{i'}_- +i\eta \psi^{i'}_+)_{\tau=0} | B_t \rangle =0}\;.
\end{array} \right.
\eea

For some special matrices, the eqs. (86)-(89) reduce to the known
cases. For example for ${\tilde A}=B={\tilde A}' = B'={\bf 0}$ and 
$A=-{\tilde B} =-A' =-{\tilde B}'=-{\bf 1}$, the directions $\{X^{\bar 
\alpha}\}$ become perpendicular to the brane. 
This is usual T-duality along the compact
directions of the brane. The case ${\tilde A}=B={\tilde A}' = B'={\bf 0}$
 and $A={\tilde B} = A'=-{\tilde B}'={\bf 1}$, is T-duality along the
 directions $\{X^{\bar i}\}$. 
 In the next section other special cases will be discussed.

{\bf Boundary state}

To know the properties of the state $\mid B_t \rangle$, we obtain this state
from the eqs. (86)-(89). In terms of oscillators, the first
equations in (86) and (88) have the form
\bea
\bigg{(} A^{\bar \alpha}_{\;\;\;{\bar \beta}} \alpha^{\bar \beta}_n  
+ B^{\bar \alpha}_{\;\;\;{\bar \beta}}\alpha^{\bar \beta}_{-n} 
+{\tilde A}^{\bar \alpha}_{\;\;\;{\bar \beta}} 
{\tilde \alpha}^{\bar \beta}_n+ 
{\tilde B}^{\bar \alpha }_{\;\;\;{\bar \beta}}
{\tilde \alpha}^{\bar \beta}_{-n} 
\bigg{)} | B_t \rangle =0 \;,  
\eea
for the bosonic part, and
\bea
\bigg{(} A^{\bar \alpha}_{\;\;\;{\bar \beta}} d^{\bar \beta}_n  
+{\tilde A}^{\bar \alpha}_{\;\;\;{\bar \beta}}{\tilde d}^{\bar \beta}_n 
-i\eta ( B^{\bar \alpha}_{\;\;\;{\bar \beta}} d^{\bar \beta}_{-n}  
+{\tilde B}^{\bar \alpha}_{\;\;\;{\bar \beta}}{\tilde d}^{\bar \beta}_{-n} 
)\bigg{)} | B_t \rangle =0 \;,  
\eea
for the R$\otimes$R sector of the
fermionic part. For the NS$\otimes$NS sector there is
\bea
\bigg{(} A^{\bar \alpha}_{\;\;\;{\bar \beta}} b^{\bar \beta}_r  
+{\tilde A}^{\bar \alpha}_{\;\;\;{\bar \beta}}{\tilde b}^{\bar \beta}_r 
-i\eta ( B^{\bar \alpha}_{\;\;\;{\bar \beta}} b^{\bar \beta}_{-r}  
+{\tilde B}^{\bar \alpha}_{\;\;\;{\bar \beta}}{\tilde b}^{\bar \beta}_{-r} 
)\bigg{)} | B_t \rangle =0 \;.  
\eea
The first equations in (87) and (89), in terms of oscillators, can be
obtained from the eqs. (90)-(92) by the changes,
\bea
\left \{ \begin{array}{cl}
\hspace{-5cm{\bar \alpha} \rightarrow {\bar i} \;\;\;,\;\;\;
{\bar \beta} \rightarrow {\bar j}} \;,\\
A \rightarrow A' \;\;,\;\; {\tilde A} \rightarrow {\tilde A}' \;\;,\;\;\;
B \rightarrow -B' \;\;\;,\;\;\; {\tilde B} \rightarrow -{\tilde B}' \;.
\end{array} \right.
\eea

Equation (90) for $n=0$, gives
\bea
\Pi^{\bar \alpha}=0 \;,
\eea
this means that in the $Y$-space closed 
string has no momentum along the $Y^{\bar \alpha}$-
direction. According to the changes (93), we also have
\bea
\Lambda^{\bar i}=0 \;\;.
\eea
In other words we obtain a relation 
between momentum (momentum numbers) and winding
numbers of closed string, around the compact
directions $\{X^{\bar \alpha}\}$ and $\{X^{\bar i}\}$,
\bea
p^{\bar \alpha} = -\frac{1}{\alpha'}f^{\bar \alpha}_{\;\;\;{\bar \beta}}
L^{\bar \beta}\;,
\eea
\bea
f=(A+B+{\tilde A}+{\tilde B})^{-1}(-A-B+{\tilde A}+{\tilde B})\;,
\eea
\bea
p^{\bar i} = -\frac{1}{\alpha'}f'^{\bar i}_{\;\;\;{\bar j}}
L^{\bar j}\;,
\eea
\bea
f'=(A'-B'+{\tilde A'}-{\tilde B'})^{-1}(-A'+B'+{\tilde A'}-{\tilde B'})\;.
\eea
In the special cases such as ${\tilde A}=B={\tilde A}' = B' ={\bf 0}$,
or $A={\tilde B}=A'={\tilde B}'= {\bf 0}$, the matrices $f$ and $f'$
are antisymmetric, therefore they can be interpreted as background
fields on the brane. For more detail of the relation between momentum and 
winding numbers of closed string see Ref. \cite{8}. 
			
As next we will concentrate on the solutions of the R$\otimes$R sector. 
The solutions of the NS$\otimes$NS sector can be obtained by similar method.
First we separate the eqs. (90) and (91) 
for positive, negative and zero values of
the integer ``$n$''. After using of the relations (48)-(53) for the matrices 
$A$, $B$, ${\tilde A}$, and ${\tilde B}$ we obtain
\bea
\bigg{(} \alpha^{\bar \alpha}_n + (A^T{\tilde B} +B^T{\tilde A})^
{\bar \alpha}_{\;\;\;{\bar \beta}} 
{\tilde \alpha}^{\bar \beta}_{-n}+(A^TB+B^TA)^{\bar \alpha}
_{\;\;\;{\bar \beta}} \alpha^{\bar \beta}_{-n} \bigg{)}
| B_t \rangle = 0 \;,
\eea
\bea
\bigg{(} {\tilde \alpha}^{\bar \alpha}_n + ({\tilde A}^T B +
{\tilde B}^T A)^
{\bar \alpha}_{\;\;\;{\bar \beta}}\alpha^{\bar \beta}_{-n} 
+({\tilde A}^T{\tilde B}
+{\tilde B}^T{\tilde A})^{\bar \alpha}
_{\;\;\;{\bar \beta}} {\tilde \alpha}^{\bar \beta}_{-n}\bigg{)}
| B_t \rangle = 0 \;,
\eea
\bea
\bigg{(} (A+B)^{\bar \alpha }_{\;\;\;{\bar \beta}}\alpha^{\bar \beta}_0
+ ({\tilde A}+{\tilde B})^{\bar \alpha }_{\;\;\;{\bar \beta}}
 {\tilde \alpha}^{\bar \beta}_0 \bigg{)}
| B_t \rangle = 0 \;,
\eea
for the bosonic part, and
\bea
\bigg{(} d^{\bar \alpha}_n -i\eta[(A^T B -B^T A)^
{\bar \alpha}_{\;\;\;{\bar \beta}}d^{\bar \beta }_{-n} 
+(A^T{\tilde B}-B^T{\tilde A})^{\bar \alpha}
_{\;\;\;{\bar \beta}} {\tilde d}^{\bar \beta}_{-n}]\bigg{)}
| B_t \rangle = 0 \;,
\eea
\bea
\bigg{(} {\tilde d}^{\bar \alpha}_n -i\eta[({\tilde A}^T 
{\tilde B} -{\tilde B}^T {\tilde A})^
{\bar \alpha}_{\;\;\;{\bar \beta}}{\tilde d}^{\bar \beta }_{-n} 
+({\tilde A}^T B-{\tilde B}^T A)^{\bar \alpha}
_{\;\;\;{\bar \beta}} d^{\bar \beta}_{-n}]\bigg{)}
| B_t \rangle = 0 \;,
\eea
\bea
\bigg{(} d^{\bar \alpha}_0 +[(A-i\eta B)^{-1}({\tilde A}-i\eta
{\tilde B})]^{\bar \alpha}_{\;\;\;{\bar \beta}} 
{\tilde d}^{\bar \beta}_0\bigg{)}
| B_t \rangle = 0 \;,
\eea
for the fermionic part. For the directions $\{ X^{\bar i}\}$, 
apply the changes (93) to the eqs. (100)-(105).

The complete boundary state in each sector is the following product
\bea
|B_t \rangle_{R,NS} = | B_t \rangle_b | B_{gh} \rangle
|B^{(f)}_t \rangle_{R,NS} | B_{sgh} \rangle_{R,NS}\;\;,
\eea
where $| B_{gh} \rangle $ and $| B_{sgh} \rangle $ are the ghost and the
superghost parts of the boundary states respectively \cite{16,17}. 
These states do not
change under the spacetime transformations (47) and (80).
The ghost part of the boundary state is
\bea
| B_{gh} \rangle = \exp \bigg{\{} \sum_{n=1}^\infty (c_{-n} {\tilde b}_{-n}
-b_{-n} {\tilde c}_{-n}) \bigg{\}} \frac{c_0 + {\tilde c}_0 }{2}
| P=1 \rangle | {\tilde P}=1 \rangle \;.
\eea
For the superghost parts there are
\bea
| B_{sgh} \rangle_{R} = \exp \bigg{\{} i\eta
\sum_{n=1}^\infty (\gamma_{-n} {\tilde \beta}_{-n}
-\beta_{-n} {\tilde \gamma}_{-n}) +i\eta \gamma_0{\tilde \beta}_0 \bigg{\}} 
| P=-1/2 \;,\; {\tilde P}= -3/2 \rangle \;,
\eea
for the R$\otimes$R sector, and
\bea
| B_{sgh} \rangle_{NS} = \exp \bigg{\{} i\eta
\sum_{r=1/2}^\infty (\gamma_{-r} {\tilde \beta}_{-r}
-\beta_{-r} {\tilde \gamma}_{-r}) \bigg{\}} 
| P=-1 \;,\; {\tilde P}= -1 \rangle \;,
\eea
for the NS$\otimes $NS sector. The ghost and the superghost vacuums
are in the $(1,1)$, $(-1/2\;,-3/2)$ and $(-1, -1)$ pictures respectively
\cite{16}.

The bosonic part of the boundary state is,
\bea
| B_t \rangle_b &=& 
\frac{T_p}{2}N_b \exp \bigg{\{} -\sum_{n=1}^{\infty}
\frac{1}{n} \bigg{[} 2\alpha^{\bar \alpha}_{-n}(A^TB)_{{\bar \alpha}
{\bar \beta}} \alpha^{\bar \beta}_{-n} + 2{\tilde \alpha}^{\bar \alpha}_{-n}
({\tilde A}^T {\tilde B})_{{\bar \alpha}{\bar \beta}}
{\tilde \alpha}^{\bar \beta}_{-n}
+\alpha^{\alpha'}_{-n}{\tilde \alpha}_{{-n}{\alpha'}}  \nonumber\\ 
&~&-\alpha^{\bar \alpha}_{-n}(A^TB)_{{\bar \alpha}
{\bar \alpha}} \alpha^{\bar \alpha}_{-n} 
-{\tilde \alpha}^{\bar \alpha}_{-n}
({\tilde A}^T {\tilde B})_{{\bar \alpha}{\bar 
\alpha}}{\tilde \alpha}^{\bar \alpha}_{-n}+  
\alpha^{\bar \alpha}_{-n}(A^T{\tilde B}+B^T{\tilde A})_{{\bar \alpha}
{\bar \beta}}{\tilde \alpha}^{\bar \beta}_{-n}\nonumber\\ 
&~&- 2\alpha^{\bar i}_{-n}(A'^TB')_{{\bar i}
{\bar j}} \alpha^{\bar j}_{-n} - 2{\tilde \alpha}^{\bar i}_{-n}
({\tilde A}'^T {\tilde B}')_{{\bar i}{\bar j}}{\tilde \alpha}^{\bar j}_{-n}
+\alpha^{\bar i}_{-n}(A'^TB')_{{\bar i}
{\bar i}} \alpha^{\bar i}_{-n} 
\nonumber\\
&~&+{\tilde \alpha}^{\bar i}_{-n}
({\tilde A}'^T {\tilde B}')_{{\bar i}{\bar 
i}}{\tilde \alpha}^{\bar i}_{-n}-  
\alpha^{\bar i}_{-n}(A'^T{\tilde B}'+B'^T{\tilde A}')_{{\bar i}
{\bar j}}{\tilde \alpha}^{\bar j}_{-n}
-\alpha^{i'}_{-n}{\tilde \alpha}_{{-n}{ i'}}
\bigg{]} \bigg{\}} | 0 \rangle\;,
\eea
where $N_b$ is an appropriate normalizing factor
and $T_p$ is the tension of the initial D$_p$-brane. 
The two terms containing the indices $\alpha'$ and $i'$, are the solutions of 
the second equations of (86) and (87), 
for the untransformed directions $\{ X^{\alpha'}\}$ and $\{ X^{i'}\}$.

The fermionic part of the R$\otimes$R sector has the solution,
\bea
| B^{(f)}_t \rangle_{R}&=& N_f \exp \bigg{\{} i\eta \sum_{n=1}^{\infty}
 \bigg{[} d^{\bar \alpha}_{-n}(A^TB)_{{\bar \alpha}
{\bar \beta}} d^{\bar \beta}_{-n} +
{\tilde d}^{\bar \alpha}_{-n}
({\tilde A}^T {\tilde B})_{{\bar \alpha}{\bar \beta}}
{\tilde d}^{\bar \beta}_{-n}
\nonumber\\ 
&~&+d^{\bar \alpha}_{-n}(A^T{\tilde B}-B^T{\tilde A})_{{\bar \alpha}
{\bar \beta}}{\tilde d}^{\bar \beta}_{-n}
- d^{\bar i}_{-n}(A'^TB')_{{\bar i}
{\bar j}} d^{\bar j}_{-n} -
{\tilde d}^{\bar i}_{-n}
({\tilde A}'^T {\tilde B}')_{{\bar i}{\bar j}}
{\tilde d}^{\bar j}_{-n}
\nonumber\\
&~&-d^{\bar i}_{-n}(A'^T{\tilde B}'-B'^T{\tilde A}')_{{\bar i}
{\bar j}}{\tilde d}^{\bar j}_{-n} 
+d^{\alpha'}_{-n}{\tilde d}_{{-n}{\alpha'}}
-d^{i'}_{-n}{\tilde d}_{{-n}{ i'}}\bigg{]} \bigg{\}} | B_t \rangle^{(0)}
_{R} \;,
\eea
where $N_f$ is a normalizing factor. The state
$\mid B_t \rangle^{(0)}_{R}$ is the solution of the zero mode part. 
The boundary state equation of 
it, is
\bea
( d^{\mu}_0 - i\eta S^{\mu}_{\;\;\;\nu}{\tilde d}^\nu_0
) | B_t \rangle^{(0)}_R = 0 \;,
\eea
where the matrix $S^\mu_{\;\;\;\nu}$ has the components,
\bea
S^{\bar \alpha}_{\;\;\;{\bar \beta}} \equiv 
Q^{\bar \alpha}_{\;\;\;{\bar \beta}} = [ (A-i\eta B)^{-1}
({\tilde B}+i\eta{\tilde A}) ] ^{\bar \alpha}_{\;\;\;{\bar \beta}}\;,
\eea
\bea
S^{\bar i}_{\;\;\;{\bar j}} \equiv 
Q'^{\bar i}_{\;\;\;{\bar j}} =[ (A'+i\eta B')^{-1}
(-{\tilde B}'+i\eta{\tilde A}')]^{\bar i}_{\;\;\;{\bar j}}\;,
\eea
\bea
S^{\alpha'}_{\;\;\;{\beta'}}=\delta^{\alpha'}_{\;\;\;{\beta'}}\;\;\;,\;\;\;
S^{i'}_{\;\;\;{j'}}=-\delta^{i'}_{\;\;\;{j'}} \;,
\eea
and the other components of the matrix
$S$ are zero. According to the conditions 
(51)-(53), both $Q$ and $Q'$ are orthogonal. 

The solution of the eq. (112) can be written as
\bea
| B_t \rangle^{(0)}_R = {\cal{M}}^{(\eta)}_{CD}
 | C \rangle \mid {\tilde D} \rangle \;,
 \eea
 where the vacuum for the fermionic zero modes 
 $d^{\mu}_0$ and ${\tilde d}^\mu_0$ is \cite{12}
 \bea
 |C \rangle | {\tilde D} \rangle = \lim_{z, {\bar z} \rightarrow 0 }
 S^C(z){\tilde S}^D({\bar z}) |0 \rangle \;,
 \eea
 in which $S^C$ and ${\tilde S}^D$ are the spin fields in the 32-dimensional
 Majorana representation.
 By substituting (116) in the eq. (112) and using the  
 Ref. \cite{12} for the action of $d^{\mu}_0$ and ${\tilde d}^\mu_0$ on the
 vacuum $ \mid C \rangle \mid {\tilde D} \rangle $, we obtain
 an equation for the $32 \times 32$-matrix ${\cal{M}}^{(\eta)}$,
\bea
(\Gamma^\mu)^T {\cal{M}}^{(\eta)} -i\eta S^\mu_{\;\;\;\nu} \Gamma_{11}
{\cal{M}}^{(\eta)} \Gamma^\nu =0 \;.
\eea
The matrix ${\cal{M}}^{(\eta)}$ can be written as 
\bea
{\cal{M}}^{(\eta)} = C' \Gamma^0 \Gamma^{a_1}...\Gamma^{a_n}
\bigg{(} \frac{1+i\eta
\Gamma_{11}}{1+i\eta}\bigg{)}G \;,
\eea
where $ C'$ is the charge conjugation matrix \cite{12}, and
the indices $\{ a_1, a_2,...,a_n \}$ are
\bea
\{ {\bar \alpha}\} \bigcup \{ {\alpha'}\} \bigcup 
\{ {\bar i}\} = \{0,a_1,...,a_n\}\;.
\eea
By separating the eq. (118) for various indices $\{\mu\}$, the matrix $G$
must satisfy the equations,
\bea
\left \{ \begin{array}{cl}
\Gamma^{i'}G=G\Gamma^{i'}\;,\\ 
\Gamma^{\alpha'}G=G\Gamma^{\alpha'} \;,\\
\Gamma^{\bar i}G=Q'^{\bar i}_{\;\;\;{\bar j}}G\Gamma^{\bar j}\;,\\ 
\Gamma^{\bar \alpha}G=Q^{\bar \alpha}_{\;\;\;{\bar \beta}}
G\Gamma^{\bar \beta}\;.
\end{array} \right.
\eea
Therefore, the matrix $G$ has the form
\bea
G= \bigg{(} e^{\frac{1}{2}{\cal{F}}_{{\bar \alpha}{\bar \beta}} 
\Gamma^{\bar \alpha}\Gamma^{\bar \beta}}\bigg{)}
\bigg{(} e^{\frac{1}{2}{\cal{F}}'_{{\bar i}{\bar j}} 
\Gamma^{\bar i}\Gamma^{\bar j}}\bigg{)} \;,
\eea
with the convention that all gamma matrices anticommute. Hence there are 
a finite number of terms in each factor. The matrices
${\cal{F}}_{{\bar \alpha}{\bar \beta}}$ and ${\cal{F'}}_{{\bar i}{\bar j}}$ 
are
\bea
{\cal{F}}= (Q+1)^{-1}(Q-1)=\bigg{(}A+{\tilde B}+i\eta({\tilde A}-B)
\bigg{)}^{-1} \bigg{(}-A+{\tilde B}+i\eta({\tilde A}+B)\bigg{)}\;,
\eea
\bea
{\cal{F}}'= (Q'+1)^{-1}(Q'-1)=\bigg{(}A'-{\tilde B}'+i\eta({\tilde A}'+B')
\bigg{)}^{-1} \bigg{(}-A'-{\tilde B}'+i\eta({\tilde A}'-B')\bigg{)}\;,
\eea
since $Q$ and $Q'$ are orthogonal, ${\cal{F}}$ and ${\cal{F}}'$ are 
antisymmetric.

The boundary state of the NS$\otimes$NS sector fermions is
\bea
| B^{(f)}_t \rangle_{NS}&=& \exp \bigg{\{} i\eta \sum_{r=1/2}^{\infty}
 \bigg{[} b^{\bar \alpha}_{-r}(A^TB)_{{\bar \alpha}
{\bar \beta}} b^{\bar \beta}_{-r} +
{\tilde b}^{\bar \alpha}_{-r}
({\tilde A}^T {\tilde B})_{{\bar \alpha}{\bar \beta}}
{\tilde b}^{\bar \beta}_{-r}
\nonumber\\
&~&+b^{\bar \alpha}_{-r}(A^T{\tilde B}-B^T{\tilde A})_{{\bar \alpha}
{\bar \beta}}{\tilde b}^{\bar \beta}_{-r} 
- b^{\bar i}_{-r}(A'^TB')_{{\bar i}
{\bar j}} b^{\bar j}_{-r} -
{\tilde b}^{\bar i}_{-r}
({\tilde A}'^T {\tilde B}')_{{\bar i}{\bar j}}
{\tilde b}^{\bar j}_{-r}
\nonumber\\ 
&~&-b^{\bar i}_{-r}(A'^T{\tilde B}'-B'^T{\tilde A}')_{{\bar i}
{\bar j}}{\tilde b}^{\bar j}_{-r}
+b^{\alpha'}_{-r}{\tilde b}_{{-r}{\alpha'}}
-b^{i'}_{-r}{\tilde b}_{{-r}{ i'}}\bigg{]} \bigg{\}} | 0 \rangle \;,
\eea

The boundary state corresponding to a D-brane \cite{12,13,14,15} 
or corresponding to a brane with internal background 
B-field \cite{2,8,9,11}, contains those closed superstring
states that the left moving and the right moving oscillator modes
couple. But the boundary states (110), (111) and (125) 
have three kinds of the
closed superstring states, which contain left-right, left-left and
right-right coupling of oscillator modes. 
In the interaction of the two transformed 
branes, these three kinds of the superstring states contribute. Therefore
these brane interactions would be 
more general than the mixed branes interactions \cite{8,9}.

\section{Special cases}
{\bf 6.1. The first special case}

Now let us study the special case 
\bea
\left \{ \begin{array}{cl}
{\tilde A}=B={\bf 0} \;, \\
{\tilde A}'=B'={\bf 0} \;.
\end{array} \right.
\eea
According to the conditions (48)-(53), the matrices $A$, $A'$, 
${\tilde B}$ and ${\tilde B}'$ are orthogonal, and the matrices $Q$,
$Q'$, ${\cal{F}}$ and ${\cal{F}}'$ are real, with ${\cal{F}}=f$ and
${\cal{F}}'=f'$. Also the boundary state equations (86)-(89) reduce
to the usual forms of the mixed brane equations.
For this case, under the transformations (47) and (80) the virasoro
operators, the super virasoro operators and BRST charge are invariant. For
example the zero mode of the super virasoro operator of the R$\otimes$R
sector
\bea
L^{(\alpha , d)}_0 = \frac{1}{4} \alpha' p^\mu p_\mu-
\frac{1}{2}p^{\bar \mu}L
_{\bar \mu} + \frac{1}{4\alpha'}L^{\bar \mu}L_{\bar \mu}
+\sum_{n=1}^{\infty}(\alpha^\mu_{-n}\alpha_{n \mu}+nd^\mu_{-n}d_{n \mu})\;,
\eea
under the transformations (47) and (80), is invariant
. Similarly ${\tilde L}^{({\tilde \alpha},{\tilde d})}_0$ is invariant. 
Therefore, the Hamiltonian of the closed 
superstring is an invariant operator. 

The normalizing factors reduce to the normalizations 
of the mixed branes \cite{9}
\bea
N_b=1/N_f&~&=\sqrt{\det(1-{\cal{F}})\det(1-{\cal{F}}')}\nonumber\\
&~& =\bigg{(} \frac{2^{k_{\bar \alpha}+k_{\bar i}}\det A \det A'}{\det(A+
{\tilde B})\det(A'-{\tilde B}')} \bigg{)}^{1/2}\;,
\eea
where $k_{\bar \alpha}$ ($k_{\bar i}$) is the 
number of the compact directions 
along (transverse to) the initial D$_p$-brane. 

The boundary states (110), (111) and (125) also take simple forms
\bea
| B_t \rangle_R =\frac{T_p}{2} \exp\bigg{\{} \sum_{n=1}^{\infty}\bigg{(}
-\frac{1}{n}\alpha^{\mu}_{-n}S_{\mu \nu}{\tilde \alpha}^{\nu}_{-n}+i\eta
d^{\mu}_{-n}S_{\mu \nu}{\tilde d}^{\nu}_{-n} \bigg{)} \bigg{\}}
| B_t \rangle^{(0)}_R \;,
\eea
\bea
| B_t \rangle_{NS} =\frac{T_p}{2} N_b 
\exp\bigg{\{} \sum_{n=1}^{\infty}\bigg{(}
-\frac{1}{n}\alpha^{\mu}_{-n}S_{\mu \nu}{\tilde \alpha}^{\nu}_{-n}
\bigg{)} +i\eta \sum_{r=1/2}^\infty \bigg{(}
b^{\mu}_{-r}S_{\mu \nu}{\tilde b}^{\nu}_{-r} \bigg{)} \bigg{\}}|0 \rangle\;,
\eea
where $S_{\mu \nu}$ is given by the 
eqs. (113)-(115), in which the eq. (126) must insert in to them.
For $\mid B_t \rangle^{(0)}_R$, insert (126) to (123) and (124) to 
obtain the matrix $G$ for this case. For appropriate matrices
${ A, A', {\tilde B} }$ and ${ {\tilde B}'}$,
the boundary states (129) and (130) describe 
a mixed brane \cite{8,9}, along the 
directions $\{ X^\alpha \} \bigcup \{ X^{\bar i}\}$, with the field strength
$({\cal{F}}^{\bar \alpha}_{\;\;\;{\bar \beta}}
\;,\; {\cal{F}}'^{\bar i}_{\;\;\;{\bar j}})$, 
whose components are on the subspaces
 $\{ X^{\bar \alpha} \}$ and $\{ X^{\bar i}\}$. Note that one can consider
 more special ${ A, A', {\tilde B} }$ and ${ {\tilde B}'}$, such that 
 some of the directions $\{X^{\bar \alpha}\}$ and $\{X^{\bar i}\}$ be 
 perpendicular to (or along) the brane. 
 For example for $A={\rm diag}(1,-1,-1,1,
 ...,1)$, two directions of $\{ X^{\bar \alpha}\}$ become perpendicular
 to the brane. For $ A={\bf -1}$ and ${\tilde B}={\bf 1}$, we have usual
 T-duality on the directions $\{X^{\bar \alpha}\}$. Similarly $A'={\bf -1}$
 and ${\tilde B}'={\bf 1}$ give T-duality on the 
 directions $\{ X^{\bar i} \}$.

{\bf 6.2. The second special case}

 The other interesting case is,

\bea
\left \{ \begin{array}{cl}
 A={ {\tilde B}}={\bf 0} \;,\\
 A' = {\tilde B}' = {\bf 0}\;.
\end{array} \right.
\eea
Now the remaining matrices $ {\tilde A}, B , {\tilde A}'$ and
$B'$ are orthogonal. Also the matrices
$Q$, $Q'$, ${\cal{F}}$ and ${\cal{F}}'$
are real. If all directions transform, i.e. 
the indices $\alpha'$ and $i'$ disappear, 
the super virasoro operators $L^{(\alpha,d)}_m$
and ${\tilde L}^{({\tilde \alpha},{\tilde d})}_m$ change to each other.
Also the exchange takes place between $F^{(\alpha,d)}_m$ and
${\tilde F}^{({\tilde \alpha},{\tilde d})}_m$, where 
\bea
F^{(\alpha, d)}_m=\sum_{n=-\infty}^{\infty} \alpha_{-n}.d_{m+n}\;,
\nonumber\\
{\tilde F}^{({\tilde \alpha}, {\tilde d})}_m
=\sum_{n=-\infty}^{\infty} {\tilde \alpha}_{-n}.{\tilde d}_{m+n}\;.
\eea

The boundary states (110), (111) and (125) take the forms,
\bea
| B_t \rangle_b &=&\frac{T_p}{2}
N_b \exp \bigg{\{} \sum_{n=1}^{\infty}\bigg{(}
-\frac{1}{n} \alpha^{\mu}_{-n}S^{(b)}_{\mu \nu}{\tilde \alpha}^{\nu}_{-n}
\bigg{)} \bigg{\}} |0 \rangle\;,
\eea
\bea
| B^{(f)}_t \rangle_R &=&N_f \exp \bigg{\{}i\eta \sum_{n=1}^{\infty} 
\bigg{(} d^{\mu}_{-n}S^{(f)}_{\mu \nu}{\tilde d}^{\nu}_{-n}
\bigg{)} \bigg{\}} | B_t \rangle^{(0)}_R \;,
\eea
\bea
| B^{(f)}_t \rangle_{NS} &=& \exp \bigg{\{} i\eta \sum_{r=1/2}^{\infty}
\bigg{(} b^{\mu}_{-r}S^{(f)}_{\mu \nu}{\tilde b}^{\nu}_{-r}
\bigg{)} \bigg{\}} | 0 \rangle \;,
\eea
where the normalizing factors are
\bea
N_b=1/N_f=\bigg{(} \frac{2^{k_{\bar \alpha}+k_{\bar i}}\det {\tilde A} 
\det {\tilde A'}}{\det({\tilde A}+
 B)\det({\tilde A'}- B')} \bigg{)}^{1/2}\;.
\eea
Also the matrices $S^{(b)}_{\mu \nu}$ and $S^{(f)}_{\mu \nu}$ are
\bea
&~&S^{\bar \alpha}_{(b)\; {\bar \beta}} = (B^T {\tilde A})^{{\bar \alpha}}
_{\;\;{\bar \beta}}\;,
\nonumber\\
&~&S^{\alpha'}_{(b)\; {\beta'}} = \delta^{\alpha'}_{\;\; \beta'}\;,
\nonumber\\
&~&S^{\bar i}_{(b)\; {\bar j}} = -(B'^T {\tilde A'})^{{\bar i}}
_{\;\;{\bar j}}\;,
\nonumber\\
&~&S^{i'}_{(b)\; {j'}} = -\delta^{i'}_{\;\; {j'}}\;,
\eea
\bea
&~&S^{\bar \alpha}_{(f)\; {\bar \beta}} = -
S^{\bar \alpha}_{(b)\; {\bar \beta}}\;,  
\nonumber\\
&~&S^{\alpha'}_{(f)\; {\beta'}} = S^{\alpha'}_{(b)\; {\beta'}}\;, 
\nonumber\\
&~&S^{\bar i}_{(f)\; {\bar j}} = -S^{\bar i}_{(b)\; {\bar j}} \;, 
\nonumber\\
&~&S^{i'}_{(f)\; {j'}} = S^{i'}_{(b)\; {j'}} \;,
\eea
the other components of $S_{(b)}$ and $S_{(f)}$ are zero. 
For zero modes state 
in (134), insert the eq. (131) to the eqs. (123) and (124) to obtain
the matrix $G$ in (122). Two matrices that appear in the
bosonic part (i.e. $S_{(b)}$) and in the fermionic part (i.e. $S_{(f)}$),
are different. For this reason we call the corresponding brane to the 
states (133)-(135) as a ``modified mixed brane''.

According to the eqs. (131) and (81), the opposite signs of 
($S^{(f)}_{{\bar \alpha} {\bar \beta} } \;,\; S^{(f)}_{{\bar i} {\bar j} }$) 
 of the fermionic part with respect to the bosonic part i.e. 
($S^{(b)}_{{\bar \alpha} {\bar \beta} } \;,\; S^{(b)}_{{\bar i} {\bar j} }$), have originated from the exchange of 
$\{d^{\bar \alpha}_n\}$ with $\{{\tilde d}^{\bar \alpha}_n\}$, 
and also $\{d^{\bar i}_n\}$
with $\{{\tilde d}^{\bar i}_n\}$. Similarly these exchanges
take place for the fermions of the NS$\otimes$NS sector.
\section{Conclusion}

We found that the $Y$-space (which 
contains the compact part of the spacetime and
its worldsheet parity transformed) 
is an interesting space for studying the 
superstring theory and its target space
dualities. For the special cases
this space reduces to the T-dual of the compact part of the spacetime,
or reduces to the T-dual of the worldsheet
parity transformed of the compact part of the spacetime. 
The $Y$-space appears similar to a compact space, 
therefore its dualities give some expected
results corresponding to a compact space.

By going from the spacetime to the $Y$-space, we obtained the 
boundary state of a closed superstring, emitted from a brane that 
is more general than the boundary state corresponding to a mixed brane. 
In other words, a brane with background NS$\otimes$NS B-field in spacetime
appears as a D-brane in $Y$-space. For example in the $Y$-space, 
closed string has no momentum along the brane, which is similar to the
closed string emitted from a wrapped mixed brane in the spacetime. 

By neglecting the worldsheet parity part of the $Y$-space, the 
transformed brane reduces to a mixed brane. 
In this case the super virasoro operators and the BRST charge are invariant.
Also by putting away the compact part of the spacetime, we obtained the
boundary state corresponding to a modified mixed brane.
The background fields of the transformed branes, 
are special combinations of the transformation matrices.

{\bf Acknowledgment}

The author would like to thank H. Arfaei for useful discussions and 
N. Sadooghi for reading the manuscript.


\end{document}